\documentclass[twocolumn,aip,nofootinbib,jcp,reprint,amsmath,amssymb]{revtex4-2}

\usepackage{natbib}
\usepackage{amsmath}
\usepackage{graphicx}
\usepackage[hidelinks]{hyperref}
\usepackage{dcolumn}
\usepackage{bm}
\usepackage{epsfig,color,xspace,multirow,xr,bbold}
\usepackage[all]{xy}
\usepackage{setspace}
\usepackage{url}
\usepackage{soul}
\usepackage[colorinlistoftodos]{todonotes}
\usepackage{longtable}
\usepackage{tabularx}
\usepackage{adjustbox}
\usepackage{threeparttable} 
\usepackage{natbib,hyperref}
\usepackage{float}
\usepackage{placeins}
\usepackage[utf8]{inputenc}
\usepackage[T1]{fontenc}
\usepackage{pifont}
\usepackage{booktabs}
\usepackage[table]{xcolor}

\definecolor{wf}{RGB}{90,90,90}

\definecolor{rsh}{RGB}{145,70,220}

\definecolor{comp}{RGB}{60,120,255}

\definecolor{dh}{RGB}{40,170,80}

\definecolor{hyb}{RGB}{245,210,40}

\definecolor{meta}{RGB}{245,140,30}

\definecolor{gga}{RGB}{220,50,45}
\definecolor{deepgreen}{RGB}{0,100,0}

\usepackage{xcolor}
\definecolor{deeppink}{RGB}{255,20,147}

\newcommand{\cs}{$C_{\rm s}$\,}
\newcommand{\ctwov}{$C_{\rm 2v}$\,}
\newcommand{\cthreeh}{$C_{\rm 3h}$\,}

\newcommand{\dthreeh}{$D_{\rm 3h}$\,}

\newcommand*{\ditto}{\texttt{"}}

\usepackage[normalem]{ulem}
\usepackage{xcolor}
\newcommand{\RRef}[1]{Ref.~\onlinecite{#1}}

\newcommand\notsotiny{\@setfontsize\notsotiny\@vipt\@viipt}
\usepackage{soul} % for highlighting
\newcommand{\orcid}[1]{\href{https://orcid.org/#1}{\includegraphics[width=8pt]{orcid.png}}}
\begin{document}

\title{
When optimized geometries are not optimal: Implications for inverted singlet–triplet gap predictions
}
\date{\today}

\author{
Atreyee Majumdar, %\orcid{0009-0007-4634-5249}
Elamathi Selvan Samynathan, %\orcid{0009-0005-2193-2602}
Raghunathan Ramakrishnan$^*$ %\orcid{0000-0002-7288-9238} 
}
\email{ramakrishnan@tifrh.res.in} 
\thanks{\\ $^*$ Corresponding author}  
\affiliation{Tata Institute of Fundamental Research, Hyderabad 500046, India}

% \keywords{
% OLEDs,
% Delayed-Fluorescence,
% Singlet-triplet energy gap,
% Heteroaromatics
% }

\begin{abstract}
\noindent 
Pseudo–Jahn–Teller distortions can substantially alter singlet–triplet gaps, potentially reversing the predicted excited-state ordering in molecules proposed to exhibit inverted singlet–triplet (INVEST) gaps. 
In this study, we benchmark geometry optimization methods for 13 azaphenalenes against CCSD(T)-quality reference geometries, assessing their ability to reproduce symmetry lowering and its consequences for singlet–triplet gaps. 
Among the methods examined, $\omega$B97M-D4 reliably describes the distorted geometries and corresponding gaps, whereas the commonly used methods B3LYP and MP2 retain high-symmetry structures, potentially leading to false-positive INVEST assignments. 
We further assess $\omega$B97M-D4 geometries across 337 molecules comprising azaphenalenes, boron/nitrogen-substituted polycyclic aromatic hydrocarbons, and non-alternant hydrocarbons. 
Subsequent L-CC2/aug-cc-pVDZ calculations predict negative singlet–triplet gaps for 85 molecules and identify discrepancies with previously reported INVEST assignments. 
These findings demonstrate that the choice of geometry optimization method is crucial for reliable INVEST screening, particularly in systems susceptible to symmetry-breaking distortions.
\end{abstract}

\maketitle

%=========================================
\section{Introduction}\label{sec:introduction}
%=========================================
Molecules exhibiting inverted singlet–triplet (INVEST) gaps, in which the lowest singlet excited state (S$_1$) lies below the lowest triplet state (T$_1$), have attracted considerable attention as potential emitters for next-generation organic light-emitting diodes (OLEDs), owing to the possibility of energetically favorable reverse intersystem crossing (rISC).\cite{aizawa2022delayed} This unusual excited-state ordering, contrary to the conventional expectation based on Hund's rule,\cite{toyota1986violation,borden1994violations} arises from electron correlation effects that can overcome the exchange contribution favoring the triplet state.\cite{dreuw2023inverted,jorner2024ultrafast,bonacic1985charge,chanda2025benchmark,majumdar2026insights} Although rare among small organic molecules,\cite{majumdar2024resilience} negative singlet–triplet gaps have been predicted in several molecular families, particularly azaphenalenes (APs),\cite{de2019inverted,ehrmaier2019singlet,loos2023heptazine,sancho2022violation,ricci2021singlet,majumdar2025leveraging} non-alternant hydrocarbons,\cite{terence2023symmetry,garner2023double,garner2024enhanced,blaskovits2024excited} and B,N-substituted polycyclic aromatic hydrocarbons.\cite{majumdar2025unlocking} The interest in APs dates back to early spectroscopic investigations of cyclazine,\cite{leupin1980low} while more recent studies have explored heptazine derivatives as promising INVEST candidates.\cite{aizawa2022delayed,sobolewski2021heptazine} Experimental evidence remains limited, although anion photoelectron spectroscopy has directly measured a negative singlet–triplet gap of $-0.047\pm0.007$ eV for pentaazaphenalene,\cite{wilson2024spectroscopic} consistent with earlier theoretical predictions.\cite{tuvckova2022origin} Additional indirect experimental evidence has been reported for pentaazaphenalene and heptazine derivatives.\cite{kusakabe2024inverted,ehrmaier2019singlet,actis2023singlet}

\begin{figure*}[!htpb]
    \centering
    \includegraphics[width=\linewidth]{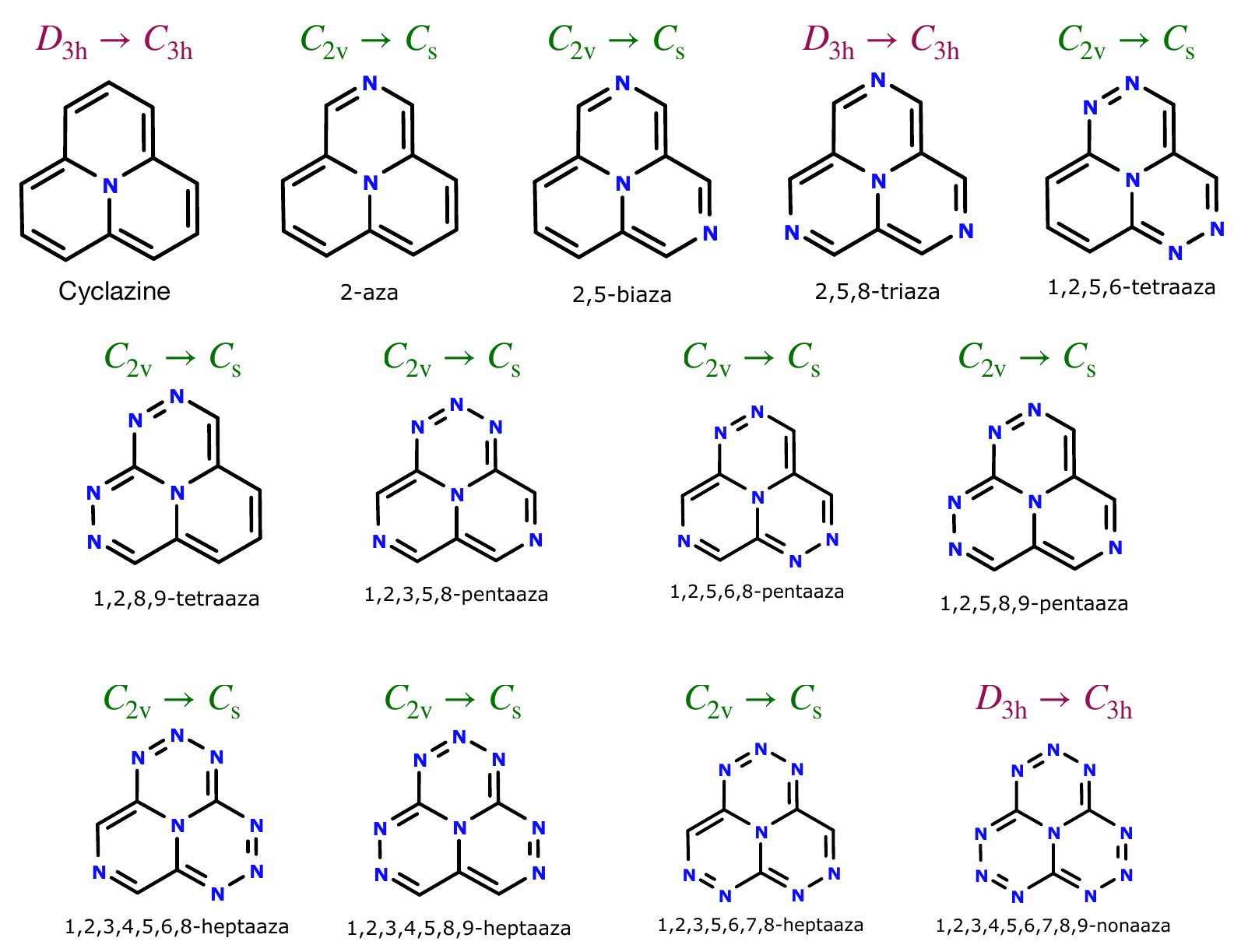}
    \caption{
Structures of the 13 distortion-prone azaphenalenes selected for benchmarking geometry optimization methods. 
Molecular names follow the convention {\it prefix}cyclazine, with the prefix indicated below each structure.  
For each molecule, the high-symmetry point group and the lower-symmetry point group resulting from structural distortion are indicated.
 }
\label{fig:13APs}
\end{figure*}

\begin{figure*}[!htpb]
    \centering
    \includegraphics[width=\linewidth]{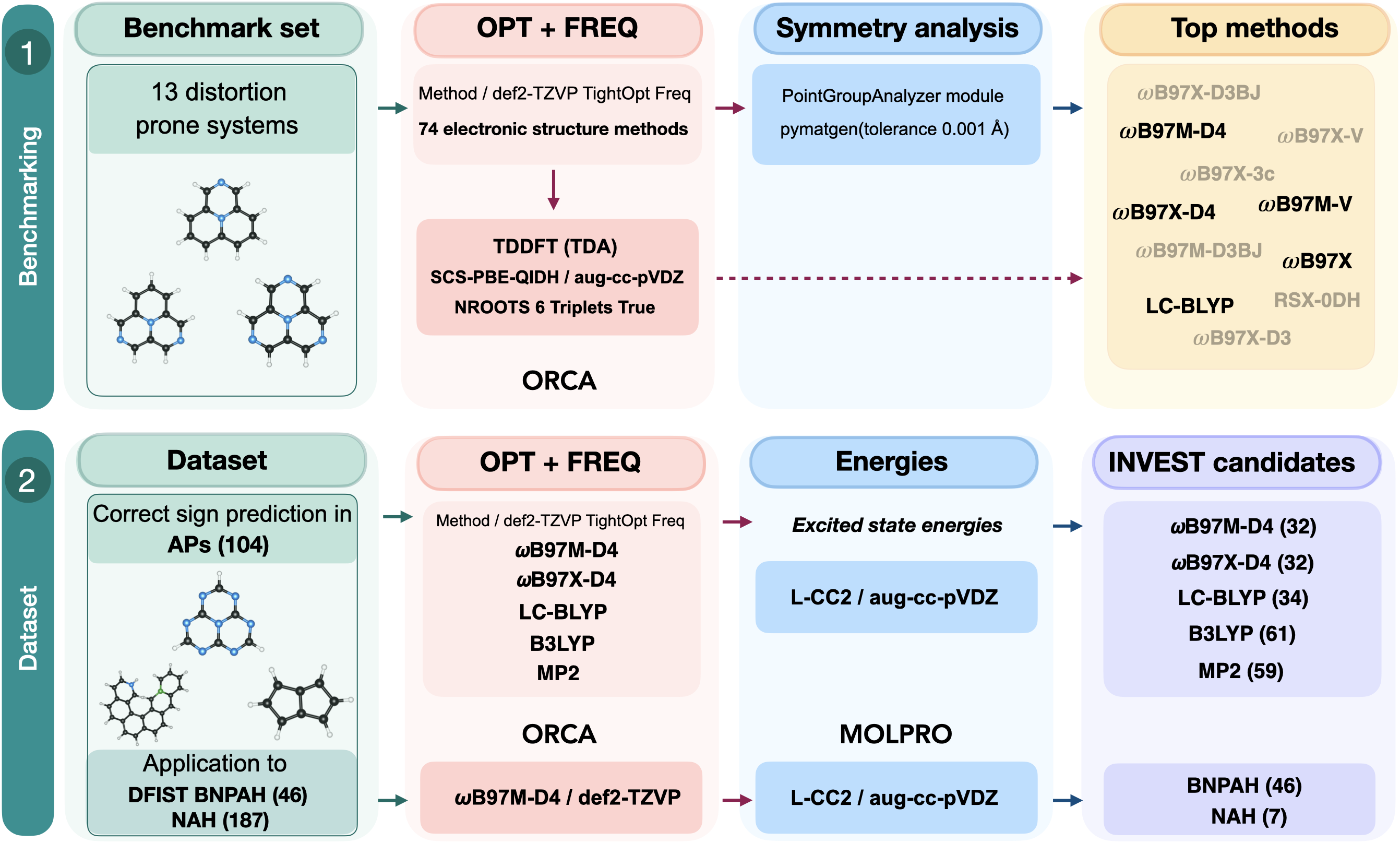}
    \caption{
Schematic overview of the two-part computational workflow. 
Part 1: Benchmarking geometry optimization methods against reference data for 13 distortion-prone azaphenalenes, focusing on molecular symmetry and singlet–triplet gaps. 
Part 2: Application of the selected geometry optimization method to three molecular datasets: APs, BN-doped polycyclic aromatic hydrocarbons (BNPAHs), and non-alternant hydrocarbons (NAHs), 
to assess geometry-dependent singlet–triplet gaps and identify potential inverted singlet–triplet (INVEST) candidates.
 }
\label{fig:workflow}
\end{figure*}

Despite extensive investigations of excited-state energies, the influence of molecular geometry on INVEST predictions has received comparatively less attention. In particular, symmetry-breaking distortions can substantially alter singlet–triplet gaps, reversing their sign from negative to positive.\cite{koseki1985violation,terence2023symmetry,majumdar2026insights,loos2023heptazine} A detailed investigation of six APs demonstrated that pseudo–Jahn–Teller distortions can significantly reduce the magnitude of negative singlet–triplet gaps or eliminate the inversion altogether.\cite{majumdar2024influence} In that study, the predicted extent of distortion depended strongly on the geometry optimization method: MP2 failed to capture the pseudo–Jahn–Teller distortion, whereas B3LYP and $\omega$B97XD provided partially satisfactory descriptions relative to CCSD(T). Therefore, high-symmetry structures may yield negative singlet–triplet gaps that do not persist upon structural relaxation, making the choice of geometry optimization method crucial for reliable INVEST assignments.

Accurately treating these distortions requires an appropriate description of the underlying potential energy surface. However, direct CCSD(T)-level geometry optimization is computationally demanding, particularly for larger conjugated molecules. A recent study employed a focal-point extrapolation scheme to obtain near-CCSD(T)/cc-pVTZ-quality geometries for 104 APs,\cite{majumdar2026insights} which have also been used to assess double-hybrid density functionals developed for INVEST systems.\cite{maiz2026sos1} Nevertheless, a computationally affordable geometry optimization method that reliably describes symmetry lowering remains desirable for larger-scale screening. Importantly, reproducing the correct point-group symmetry alone does not guarantee structural accuracy, since a method may capture the distortion qualitatively while overestimating or underestimating its magnitude. Such differences can influence the calculated excitation energies and 
ultimately, the sign of the predicted singlet–triplet gap.

In this work, we investigate the performance of geometry optimization methods for 13 pseudo–Jahn–Teller distortion-prone APs with CCSD(T)-level reference geometries (Fig.~\ref{fig:13APs}). We assess the methods by their ability to reproduce symmetry lowering and the resulting S$_1$ and T$_1$ excitation energies and singlet–triplet gaps, calculated consistently at the SCS-PBE-QIDH/aug-cc-pVDZ level. Based on this assessment, we select $\omega$B97M-D4 for a broader investigation of 337 molecules comprising 104 APs, 46 B,N-substituted polycyclic aromatic hydrocarbons, and 187 non-alternant hydrocarbons. Using $\omega$B97M-D4-optimized geometries, we perform a consistent set of L-CC2/aug-cc-pVDZ excited-state calculations to examine the sensitivity of singlet–triplet gaps to molecular geometry and reassess INVEST predictions across the three datasets.

\begin{table*}[htbp]
\centering
\caption{
Distribution of point-group symmetries obtained from geometry optimizations of 13 distortion-prone azaphenalenes using 74 computational methods. 
Each row lists the number of molecules assigned to the \cthreeh, \dthreeh, \cs, and \ctwov point groups, together with the corresponding methods. 
The predicted symmetry alone does not establish structural accuracy, as some methods may overestimate the magnitude of the distortions.
}
\begin{tabular}{c c c c l}
\hline
 \cthreeh & \dthreeh &  \cs & \ctwov & Methods \\
\hline
 3  &  0  &  10  & 0       &  CCSD(T), $\omega$B97, $\omega$B97M-V, $\omega$B97M-D4, $\omega$B97M-D3BJ, $\omega$B97X, $\omega$B97X-V, $\omega$B97X-D4,
  \\         
&&&&
$\omega$B97X-D3BJ,
%HF-3c,
 $\omega$B97X-3c,
LC-PBE,
LC-BLYP,
RSX-0DH
\\
\hline

3 & 0 & 9 & 1 &
$\omega$Pr2SCAN50,
HF
\\
\hline

2 & 1 & 9 & 1 &
$\omega$B97X-D3,
Pr2SCAN69,
revDSD-PBEP86/2021,
revDSD-PBEP86-D4/2021
\\
\hline

2 & 1 & 10 & 0 &
revDOD-PBEP86-D4/2021,
revDOD-PBEP86/2021
\\
\hline

2 & 1 & 8 & 2 &
CAM-B3LYP,
$\omega$B2PLYP
\\
\hline

1 & 1 & 10 & 1 &
M06-2X
\\
\hline

1 & 2 & 8 & 2 &
BHANDLYP,
RSX-QIDH
\\
\hline

1 & 2 & 7 & 3 &
$\omega$B2GP-PLYP,
$\omega$B97X-2
\\
\hline

1 & 2 & 6 & 4 &
Pr2SCAN50,
$\omega$B97XD
\\
\hline

1 & 2 & 4 & 6 &
PBEh-3c,
mPW1LYP,
B2PLYP-D4
\\
\hline

1 & 2 & 3 & 7 &
B3LYP,
B2K-PLYP,
B2PLYP,
DSD-PBEP86,
PBE0-DH,
mPW2PLYP,
$\omega$PBEPP86
\\
\hline

1 & 2 & 2 & 8 &
mPW1PW,
B2GP-PLYP,
PBE-QIDH
\\
\hline

1 & 2 & 1 & 9 &
B3LYP/G,
PBE0,
revPBE0,
X3LYP,
TPSS0,
TPSSh,
r2SCAN0,
B2T-PLYP
\\
\hline

0 & 2 & 5 & 6 &
M06
\\
\hline

0 & 3 & 4 & 6 &
M06-L,
BLYP,
DSD-BLYP
\\
\hline

0 & 3 & 3 & 7 &
B97M-D3BJ,
PBE,
revPBE,
mPWLYP,
mPWPW,
TPSS
\\
\hline

0 & 3 & 2 & 8 &
B97M-D4,
B97M-V,
B97-3c,
BP86,
PW91,
r2SCANh
\\
\hline

0 & 3 & 1 & 9 &
VWN5,
R2SCAN-3c,
XLYP,
r2SCAN
\\
\hline

1 & 2 & 5 & 5 &
$\omega$B88PP86
\\
\hline

1 & 2 & 0 & 10 &
MP2
\\
\hline

\end{tabular}
\vspace{2mm}

\label{tab:method_symm}
\end{table*}

%=========================================
\section{Computational Details}\label{sec:compdetails}
%=========================================

\subsection{Geometry optimization benchmark}

We selected thirteen APs (see Fig.~\ref{fig:13APs}) exhibiting pseudo--Jahn--Teller distortions from \RRef{majumdar2026insights} to assess the performance of different electronic structure methods in predicting symmetry lowering and the resulting molecular geometries.  
The reference structures, obtained using the extrapolated CCSD(T)/cc-pVTZ scheme reported in \RRef{majumdar2026insights}, were used as initial geometries for subsequent optimizations with density functional theory (DFT), double-hybrid DFT (DH-DFT), Hartree--Fock (HF), and second-order M{\o}ller--Plesset perturbation theory (MP2).  

The density functional approximations considered span all five rungs of Jacob's ladder.\cite{perdew2001jacob} These include (i) the local density approximation (LDA), represented by VWN5;\cite{wilk1981fermi} (ii) generalized gradient approximations (GGAs), comprising BP86, BLYP, XLYP, PW91,\cite{perdew1992accurate,burke1998derivation} PBE,\cite{perdew1996generalized} mPWPW, mPWLYP, and revPBE; and (iii) meta-GGAs, including TPSS,\cite{perdew2009workhorse} M06-L,\cite{zhao2006new} B97M-V, B97M-D3BJ, B97M-D4, and r$^2$SCAN.

The fourth rung comprises global and range-separated hybrid functionals. The global hybrid GGAs considered were B3LYP,\cite{stephens1994ab} B3LYP/G, PBE0,\cite{adamo1999toward} revPBE0, X3LYP,\cite{su2004accurate} mPW1PW, mPW1LYP, and BHandHLYP.\cite{becke1993new} The global hybrid meta-GGAs included TPSSh, TPSS0,\cite{perdew2009workhorse} M06, M06-2X,\cite{zhao2008m06} r$^2$SCANh, and r$^2$SCAN0. The range-separated hybrid GGAs comprised $\omega$B97 and $\omega$B97X,\cite{chai2008systematic} $\omega$B97X-V, $\omega$B97X-D3, $\omega$B97X-D3BJ, $\omega$B97X-D4, $\omega$B97XD, CAM-B3LYP,\cite{yanai2004new} LC-BLYP,\cite{tawada2004long} and LC-PBE.\cite{iikura2001long} The range-separated hybrid meta-GGAs included $\omega$B97M-V, $\omega$B97M-D3BJ, and $\omega$B97M-D4.

The fifth rung comprises double-hybrid functionals, which incorporate perturbative correlation in addition to density-functional contributions. The global double hybrids considered were B2PLYP,\cite{grimme2006semiempirical} B2PLYP-D4, mPW2PLYP,\cite{schwabe2006towards} B2GP-PLYP,\cite{karton2008highly} B2K-PLYP and B2T-PLYP,\cite{tarnopolsky2008double} PBE-QIDH,\cite{bremond2014communication} PBE0-DH,\cite{bremond2011seeking} DSD-BLYP and DSD-PBEP86,\cite{yu2014spin} revDSD-PBEP86/2021, revDSD-PBEP86-D4/2021, revDOD-PBEP86/2021, revDOD-PBEP86-D4/2021, Pr2SCAN50, and Pr2SCAN69. The range-separated double hybrids included $\omega$B97X-2,\cite{chai2009long} $\omega$B2PLYP and $\omega$B2GP-PLYP,\cite{casanova2019omegab2plyp} $\omega$B88PP86 and $\omega$PBEPP86,\cite{casanova2021time} RSX-QIDH,\cite{bremond2018range} and RSX-0DH.\cite{bremond2019range} 
We also examined the composite methods B97-3c, PBEh-3c, r$^2$SCAN-3c, and $\omega$B97X-3c.

We performed geometry optimizations with the def2-TZVP basis set, except for the composite methods, which use their respective predefined basis sets. 
The $\omega$B97XD and MP2 optimizations were carried out using Gaussian 16 (revision C.01),\cite{frisch2016gaussianshort} while the remaining optimizations were performed using ORCA (version 6.0.0).\cite{neese2012orca,neese2018software,neese2025software} 

During geometry optimizations, we did not impose symmetry constraints and used harmonic vibrational frequencies to verify that the optimized structures corresponded to local minima. 
We assessed the performance of the geometry optimization methods in two stages. 
First, we determined the point-group symmetries of the optimized structures using the \texttt{PointGroupAnalyzer} module of \texttt{Pymatgen}~\cite{ong2013python} (tolerance: 0.001~\AA) to assess whether the methods reproduced the symmetry lowering observed at the CCSD(T) level.

As predicting a lower-symmetry structure does not necessarily imply an accurate description of the distortion, a second assessment examined how the optimized geometries affected the excited-state energies as follows. For each geometry, we calculated the S$_1$ and T$_1$ excitation energies and the corresponding singlet–triplet gaps using SCS-PBE-QIDH/aug-cc-pVDZ within the Tamm--Dancoff approximation (TDA). We applied the same excited-state method to the CCSD(T)-level reference geometries, allowing us to evaluate geometry-optimization errors consistently. SCS-PBE-QIDH is a double-hybrid approach that has performed well for singlet–triplet gaps in related conjugated molecular systems.\cite{sancho2022violation,loos2023heptazine,majumdar2024resilience} All TDA calculations were performed using ORCA (version 6.0.0).\cite{neese2012orca,neese2018software}

\subsection{Application to molecular datasets}

Based on the geometry optimization benchmark, we selected $\omega$B97M-D4 for a broader assessment of 337 molecules from three datasets: 104 azaphenalenes,\cite{majumdar2026insights} 46 B,N-doped polycyclic aromatic hydrocarbons (BNPAHs),\cite{majumdar2025unlocking} and 187 non-alternant hydrocarbons (NAHs).\cite{garner2023double} All 337 molecular geometries were reoptimized at the $\omega$B97M-D4/def2-TZVP level.

Using these optimized geometries, we calculated S$_1$ and T$_1$ excitation energies and the corresponding singlet–triplet gaps consistently at the L-CC2/aug-cc-pVDZ level. L-CC2 employs a Laplace-transform-based treatment of the doubles contributions and has previously been shown to provide a favorable balance between accuracy and computational cost for excited-state calculations of APs.\cite{majumdar2025leveraging} 
For the singlet–triplet gaps of twelve triangulene molecules studied in \RRef{majumdar2025leveraging}, L-CC2 and L-ADC(2) with the aug-cc-pVTZ basis set yielded mean absolute errors (MAEs) of $0.015$ and $0.022$ eV, respectively, while CC2 and ADC(2) yielded MAEs of $0.013$ and $0.019$ eV, respectively~\cite{loos2023heptazine,loos2025correction} with respect to a composite theoretical best estimate scheme~\cite{loos2023heptazine}. 
With the aug-cc-pVDZ basis set, the corresponding MAEs of L-CC2 and L-ADC(2) were $0.016$ and $0.023$ eV, respectively, while offering substantial computational savings.
We compare the resulting singlet–triplet gaps with available literature data to assess the sensitivity of INVEST predictions to the geometry optimization method. All L-CC2 calculations were performed using Molpro (version 2015.1).\cite{werner2015molpro}

\begin{table*}[htbp]
\centering
\caption{
Error statistics for S$_1$ and T$_1$ excitation energies and singlet–triplet gaps calculated at the SCS-PBE-QIDH/aug-cc-pVDZ level using geometries optimized with different methods, relative to the corresponding values obtained at CCSD(T)-optimized geometries. 
The methods were selected based on their performance in predicting molecular symmetry as per  Table~\ref{tab:method_symm}). 
Mean absolute errors (MAEs) and standard deviations of errors (SDEs) are reported for each quantity and collectively (all). All errors are in eV.
}
\label{tab:mae_s1t1}
\begin{tabular}{lcccc c cccc}
\hline
Method & \multicolumn{4}{l}{MAE} && \multicolumn{4}{l}{SDE} \\
\cline{2-5} \cline{7-10} 
      & S$_1$ & T$_1$ & gap & all && S$_1$ & T$_1$ & gap & all \\
\hline
$\omega$B97M-D4          & 0.065 & 0.042 & 0.034 & 0.047 && 0.074 & 0.048 & 0.030 & 0.058 \\
$\omega$B97M-V           & 0.067 & 0.041 & 0.039 & 0.049 && 0.077 & 0.048 & 0.032 & 0.060 \\
LC-BLYP                  & 0.070 & 0.043 & 0.044 & 0.052 && 0.083 & 0.050 & 0.035 & 0.064 \\
$\omega$B97X             & 0.075 & 0.058 & 0.030 & 0.054 && 0.083 & 0.057 & 0.031 & 0.066 \\
$\omega$B97X-D4          & 0.076 & 0.057 & 0.030 & 0.054 && 0.083 & 0.054 & 0.031 & 0.065 \\
$\omega$B97X-D3BJ        & 0.076 & 0.056 & 0.030 & 0.054 && 0.083 & 0.054 & 0.032 & 0.065 \\
$\omega$B97X-V           & 0.078 & 0.054 & 0.032 & 0.054 && 0.086 & 0.054 & 0.034 & 0.066 \\
$\omega$B97M-D3BJ        & 0.081 & 0.049 & 0.043 & 0.058 && 0.103 & 0.058 & 0.050 & 0.077 \\
 $\omega$B97X-3c         & 0.083 & 0.060 & 0.034 & 0.059 && 0.095 & 0.064 & 0.035 & 0.074 \\
RSX-0DH                  & 0.092 & 0.070 & 0.053 & 0.072 && 0.108 & 0.083 & 0.040 & 0.090 \\
$\omega$B97X-D3          & 0.113 & 0.040 & 0.081 & 0.078 && 0.100 & 0.050 & 0.051 & 0.079 \\
revDOD-PBEP86/2021       & 0.122 & 0.057 & 0.066 & 0.081 && 0.021 & 0.013 & 0.015 & 0.033 \\
revDOD-PBEP86-D4/2021    & 0.126 & 0.058 & 0.068 & 0.084 && 0.021 & 0.013 & 0.015 & 0.034 \\
Pr2SCAN69                & 0.137 & 0.056 & 0.081 & 0.091 && 0.029 & 0.014 & 0.018 & 0.040 \\
$\omega$B2PLYP           & 0.139 & 0.047 & 0.100 & 0.095 && 0.100 & 0.050 & 0.052 & 0.082 \\
$\omega$B97              & 0.154 & 0.121 & 0.043 & 0.106 && 0.066 & 0.064 & 0.019 & 0.070 \\
LC-PBE                   & 0.159 & 0.150 & 0.035 & 0.115 && 0.073 & 0.089 & 0.035 & 0.088 \\
CAM-B3LYP                & 0.172 & 0.058 & 0.118 & 0.116 && 0.101 & 0.046 & 0.058 & 0.087 \\
revDSD-PBEP86/2021       & 0.176 & 0.081 & 0.094 & 0.117 && 0.028 & 0.021 & 0.021 & 0.048 \\
revDSD-PBEP86-D4/2021    & 0.179 & 0.083 & 0.096 & 0.119 && 0.029 & 0.021 & 0.022 & 0.049 \\
M06-2X                    & 0.180 & 0.060 & 0.120 & 0.120 && 0.068 & 0.033 & 0.038 & 0.069 \\
$\omega$B97XD            & 0.186 & 0.066 & 0.120 & 0.124 && 0.083 & 0.041 & 0.043 & 0.077 \\
B3LYP                    & 0.360 & 0.152 & 0.208 & 0.240 && 0.116 & 0.060 & 0.075 & 0.124 \\
MP2                      & 0.415 & 0.194 & 0.221 & 0.277 && 0.163 & 0.088 & 0.096 & 0.156 \\
HF                       & 0.447 & 0.303 & 0.170 & 0.307 && 0.184 & 0.169 & 0.125 & 0.198 \\
$\omega$Pr2SCAN50        & 0.460 & 0.322 & 0.183 & 0.322 && 0.418 & 0.306 & 0.165 & 0.321 \\
% HF-3c                   & 0.647 & 0.402 & 0.245 & 0.431 & 0.077 & 0.094 & 0.117 & 0.192 \\

\hline

\hline
\end{tabular}
\vspace{2mm}

\label{tab:method_error_singlet–triplet gap}
\end{table*}

%=========================================
\section{Results and Discussion}\label{sec:results}
%=========================================

We present the results in two parts. 
In the first part, we assess the performance of 74 geometry optimization methods for 13 pseudo–Jahn--Teller  distortion-prone azaphenalenes, focusing on their ability to reproduce the reference point-group symmetries and the resulting excited-state energies and singlet–triplet gaps. 
Based on these assessments, we select the most reliable geometry optimization method for further investigation. 
In the second part, we apply the selected method to three molecular datasets and compare the singlet–triplet gaps calculated from the reoptimized geometries with previously reported values to assess how geometry optimization affects INVEST predictions. 
Fig.~\ref{fig:workflow} illustrates the overall workflow.

\subsection{Benchmarking of optimization methods\label{label:subsec1}}

We first examined the ability of different electronic structure methods to reproduce symmetry-breaking distortions in 13 APs, for which extrapolated CCSD(T)-level calculations predict lower-symmetry minima comprising three \cthreeh and ten \cs structures.\cite{majumdar2026insights} 

Table~\ref{tab:method_symm} summarizes the distribution of the optimized structures among four point groups (\cthreeh, \dthreeh, \ctwov, and \cs). Thirteen methods reproduce the same point-group distribution as the CCSD(T)-level reference, with range-separated hybrid functionals accounting for the majority of these methods. This trend indicates the
need for appropriate treatment of exchange interaction at various length scales lacking in global hybrid DFT approximations. 
In contrast, HF predicts one of the ten reference \cs structures to have \ctwov symmetry, while MP2 retains \ctwov symmetry for all ten. 
These results highlight the strong dependence of symmetry-breaking distortions on the geometry optimization method. 
%However, reproducing the reference point-group distribution does not necessarily ensure structural accuracy, since the distortion magnitude may still be overestimated or underestimated.

\begin{table*}[!htpb]
\centering
\caption{
Summary of the three molecular datasets investigated in this study, showing the geometry optimization and excited-state calculation methods employed in the original studies and the present work. The corresponding literature references and numbers of molecules predicted to exhibit inverted singlet–triplet (INVEST) gaps are also provided.
}
\label{tab:exact}
\begin{tabular}{l ll c c}
\toprule
Dataset & \multicolumn{2}{l}{Method} & Source & INVEST candidates \\
\cmidrule(lr){2-3}
 & Geometry optimization & Excited states &  &  \\
\midrule
 AP-104 & ${\rm CCSD(T)}$  & L-CC2/aug-cc-pVDZ  & \RRef{majumdar2026insights} & 30 \\
 \ditto & $\omega$B97M-D4/def2-TZVP  & L-CC2/aug-cc-pVDZ  & This work & 32 \\
 DFIST-BNPAH & $\omega$B97X-D3/def2-TZVP & L-CC2/aug-cc-pVDZ & \RRef{majumdar2025unlocking} & 46  \\
 \ditto & $\omega$B97M-D4/def2-TZVP  & L-CC2/aug-cc-pVDZ  & This work & 46 \\
 187 non-alternant hydrocarbons & $\omega$B97XD/def2-TZVP &  CC2/aug-cc-pVDZ & \RRef{garner2023double} & 8 \\
  \ditto & $\omega$B97M-D4/def2-TZVP &  L-CC2/aug-cc-pVDZ & This work & 7 \\
\bottomrule
\end{tabular}
\label{tab:datasetinfo}
\end{table*}

To assess the effects on excited-state properties, we calculated S$_1$ and T$_1$ excitation energies and the corresponding singlet--triplet gaps at the SCS-PBE-QIDH/aug-cc-pVDZ level using geometries obtained with different optimization methods. SCS-PBE-QIDH has previously demonstrated good performance for singlet–triplet gap predictions in related molecular systems.\cite{majumdar2024resilience,loos2023heptazine} We compared the results with those calculated at the same excited-state level using CCSD(T)-level reference geometries, thereby isolating errors arising from geometry optimization. Mean absolute errors (MAEs) and standard deviations of errors (SDEs) were evaluated separately for S$_1$, T$_1$, and singlet–triplet gaps, as well as collectively for all three quantities. Table~\ref{tab:method_error_singlet–triplet gap} presents the corresponding statistics for the selected methods.

Among the methods examined, $\omega$B97M-D4 provides the lowest overall MAE of 0.047 eV, with an MAE of 0.034 eV for singlet–triplet gaps. The closely related $\omega$B97M-V functional also performs well, with an overall MAE of 0.049 eV. However, converged geometries with $\omega$B97M-V required numerical frequencies, increasing computational cost and making it less attractive for large-scale applications. Other methods showing relatively low errors include LC-BLYP, $\omega$B97X, and $\omega$B97X-D4, with overall MAEs of 0.052, 0.054, and 0.054 eV, respectively. In comparison, $\omega$B97X-D3, used in previous INVEST studies, shows a larger overall MAE of 0.078 eV. B3LYP and MP2 yield substantially larger errors, with overall MAEs of 0.240 and 0.277 eV, respectively. 
In all, these results identify $\omega$B97M-D4 as a promising choice for describing pseudo–Jahn–Teller-distorted geometries while minimizing geometry-induced errors in excited-state energies.

To further examine the effect of geometry optimization on the qualitative prediction of INVEST behavior, we compared the S$_1$ and T$_1$ excitation energies and singlet–triplet gaps calculated at the L-CC2/aug-cc-pVDZ level using geometries optimized with $\omega$B97M-D4, $\omega$B97X-D4, B3LYP, and MP2 (Table S1 of the Supplementary Information) with those obtained using CCSD(T)-level reference geometries. 
Both $\omega$B97M-D4 and $\omega$B97X-D4 reproduce the reference singlet–triplet gap signs for 12 of the 13 molecules. The exception is 1,2,8,9-tetraazacyclazine, for which the singlet–triplet gaps are $-0.051$ and $-0.026$ eV at the $\omega$B97M-D4 and $\omega$B97X-D4 geometries, respectively, compared with $+0.009$ eV at the reference geometry. In contrast, L-CC2 calculations using B3LYP- and MP2-optimized geometries predict negative singlet–triplet gaps for all systems except 1,2,3,4,5,6,7,8,9-nonaazacyclazine (10AP). Thus, B3LYP and MP2 may inadequately describe symmetry-breaking distortions, leading to false-positive INVEST assignments. These findings show that the reliability of a geometry optimization method must be assessed not only by its predicted molecular symmetry but also by the resulting excited-state properties.

\begin{figure}[!htpb]
    \centering
    \includegraphics[width=\linewidth]{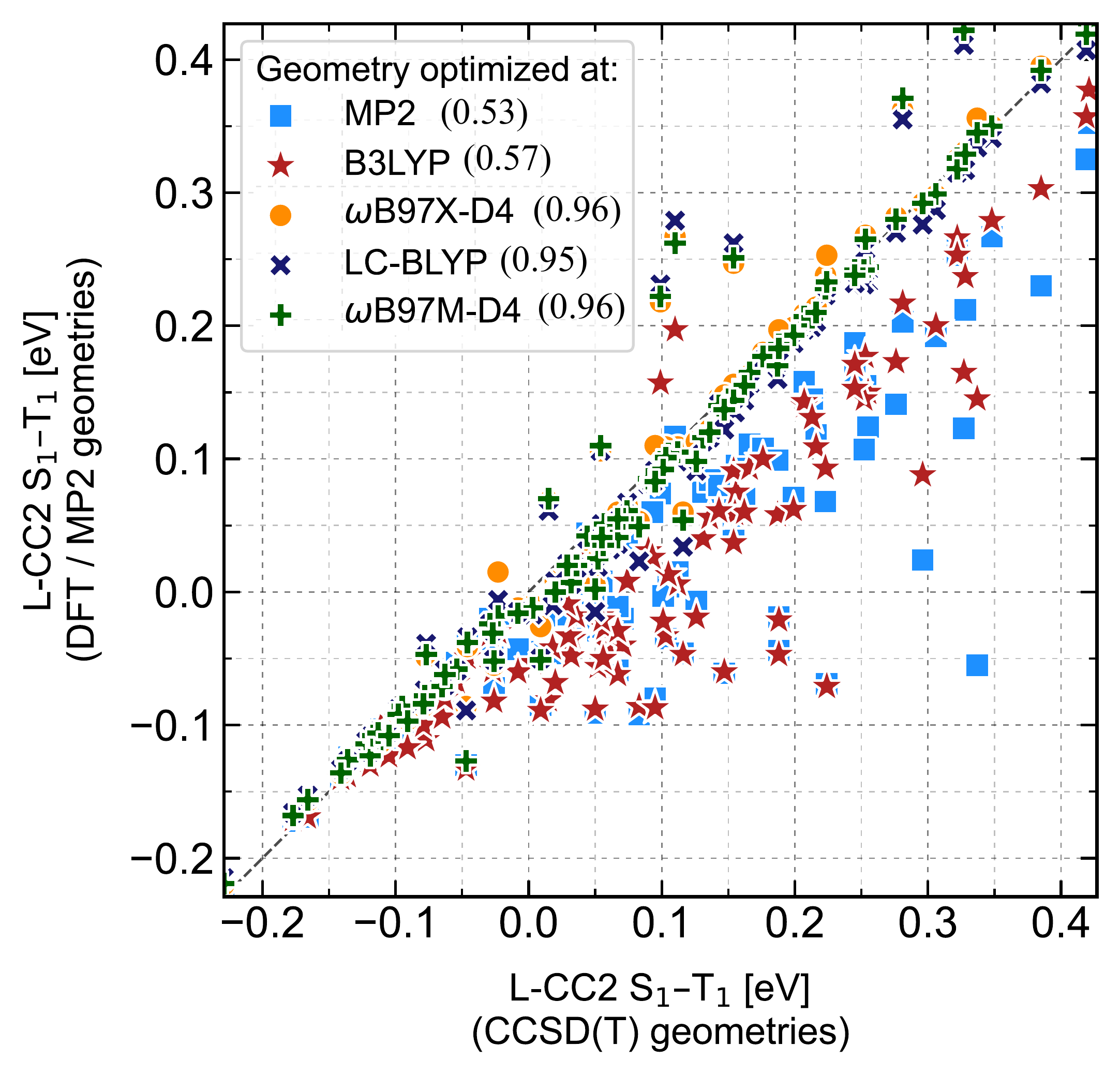}
\caption{
Comparison of singlet–triplet gaps for 104 azaphenalenes computed at geometries optimized with $\omega$B97M-D4, $\omega$B97X-D4, B3LYP, and MP2, compared with those obtained at CCSD(T)-optimized geometries. 
The coefficient of determination ($R^2$) for each method is given in parentheses.
}
\label{fig:scatter}
\end{figure}

\subsection{Singlet-triplet gaps based on $\omega$B97M-D4 geometries using diverse Datasets\label{label:dataset}}

We now aim to verify whether the conclusions obtained from the chosen methods remain valid for a broader range of systems. Hence, we have selected 3 sets of systems for this analysis. 
The selected datasets consist of three system types, and all contain molecules exhibiting the INVEST property, as shown in previous studies. 
The three datasets include 104 azaphenalenes, 46 B,N-substituted polyaromatic hydrocarbons, and 187 nonalternant hydrocarbons. 
 Table~\ref{tab:datasetinfo} summarizes the three benchmark datasets and compares the number of INVEST molecules reported in the original studies with those obtained in the present work. 
 Although the literature classifications used different geometry optimization and excited-state methods, we re-evaluated all molecules using the optimized geometries from our benchmarking, along with L-CC2 excitation-energy calculations, to consistently reassess the reported INVEST candidates.

\subsubsection{AP-104}

The first dataset comprises 104 APs selected from the AP117 dataset.\cite{AP117} 
Previous calculations using extrapolated CCSD(T)-level geometries and L-CC2/aug-cc-pVDZ excitation energies identified 30 molecules with negative singlet--triplet gaps.\cite{majumdar2026insights} 
These results provide a reference for assessing how geometry-optimization methods affect INVEST predictions. We considered five methods: $\omega$B97M-D4, $\omega$B97X-D4, and LC-BLYP, which performed well in the initial benchmark, together with B3LYP and MP2, which showed substantial deficiencies in describing symmetry-breaking distortions (Section~\ref{label:subsec1}). 
For each optimized geometry, we calculated the S$_1$ and T$_1$ excitation energies and corresponding singlet–triplet gaps consistently at the L-CC2/aug-cc-pVDZ level.

 Figure~\ref{fig:scatter} compares the calculated singlet–triplet gaps with those from CCSD(T)-level reference geometries, while Table~\ref{tab:error_method_comparison} summarizes the statistical errors and sign-classification performance. 
 The three best-performing methods agree closely with the reference values, with mean absolute errors (MAEs) of 0.0170, 0.0159, and 0.0198 eV for $\omega$B97M-D4, $\omega$B97X-D4, and LC-BLYP, respectively. 
 The corresponding standard deviations of errors (SDEs) are 0.0310, 0.0290, and 0.0330 eV. Among these methods, $\omega$B97X-D4 gives the lowest MAE and SDE and a coefficient of determination ($R^2$) of 0.9607. However, the differences in numerical accuracy are relatively small.

 For INVEST screening, accurately predicting the sign of the singlet–triplet gap is particularly important, since even a small geometry-induced error can change the qualitative excited-state ordering. 
 When we take the 30 molecules with negative reference singlet–triplet gaps as positive cases, $\omega$B97M-D4 correctly identifies all 30, with only two false positives. 
 LC-BLYP also identifies all 30 reference positives but produces four false positives. In comparison, $\omega$B97X-D4 gives 29 true positives, three false positives, and one false negative. 
 Thus, although $\omega$B97X-D4 yields marginally smaller statistical errors, $\omega$B97M-D4 provides better sign-classification performance.

  \begin{table}[htbp]
\centering
\caption{
Statistical errors and singlet–triplet gap sign-classification performance of five geometry optimization methods for 104 azaphenalenes, relative to results obtained using extrapolated CCSD(T)-level reference geometries. We calculated all excited-state energies at the L-CC2/aug-cc-pVDZ level. Mean signed errors (MSEs), mean absolute errors (MAEs), and standard deviations of errors (SDEs) are reported in eV. Sign-classification performance is quantified using true positives (TP), true negatives (TN), false positives (FP), and false negatives (FN), with negative singlet–triplet gaps classified as positive cases.
}
\label{tab:error_method_comparison}
\small
\setlength{\tabcolsep}{4pt}
\renewcommand{\arraystretch}{1.2}
\begin{tabular}{lrrrrrrr}
\hline
Method & MSE & MAE & SDE & TP & TN & FP & FN \\
\hline
$\omega$B97M-D4 & $-0.0003$ & $0.0170$ & $0.0310$ & $30$ & $72$ & $2$ & $0$  \\
LC-BLYP & $-0.0030$ & $0.0198$ & $0.0330$ & $30$ & $70$ & $4$ & $0$  \\
$\omega$B97X-D4 & $0.0028$ & $0.0159$ & $0.0290$ & $29$ & $71$ & $3$ & $1$ \\
B3LYP & $-0.0738$ & $0.0776$ & $0.0622$ & $30$ & $43$ & $31$ & $0$  \\
MP2 & $-0.0719$ & $0.0738$ & $0.0713$ & $30$ & $45$ & $29$ & $0$ \\
\hline
\end{tabular}
\end{table}

 In contrast, B3LYP and MP2 exhibit substantially larger errors and a systematic bias toward more negative singlet–triplet gaps. 
 Their mean signed errors (MSEs) are $-0.0738$ and $-0.0719$ eV, respectively, with corresponding MAEs of 0.0776 and 0.0738 eV. 
 Although both methods recover all 30 reference negative-gap systems, they incorrectly predict negative singlet–triplet gaps for an additional 31 and 29 molecules (i.e., false positive INVEST predictions), respectively. 
 Their lower $R^2$ values of 0.57 and 0.53 further indicate worse agreement with the reference data. 
 These results demonstrate that geometry optimization methods that inadequately describe symmetry-breaking distortions can substantially overestimate the number of predicted INVEST candidates.

A detailed comparison of molecules exhibiting negative singlet–triplet gaps at the reference or selected DFT geometries is provided in Table S2 of the Supplementary Information. 
While 30 APs exhibit negative singlet–triplet gaps at the CCSD(T)-level reference geometries, both $\omega$B97M-D4 and $\omega$B97X-D4 predict 32 negative-gap systems. 
However, these two DFT methods predict different sets of molecules. 
The $\omega$B97M-D4 geometries retain all 30 reference negative-gap molecules and yield two additional negative-gap predictions. In contrast, $\omega$B97X-D4 misses one reference negative-gap molecule, 1,3,5,8-tetraazacyclazine, while introducing three false positives. Notably, 1,2,8,9-tetraazacyclazine, one of the distortion-prone systems examined in the initial benchmark, is predicted to have a negative singlet–triplet gap at both DFT-optimized geometries despite its positive reference gap.

Overall, $\omega$B97M-D4 offers the most favorable balance between numerical accuracy and reliable singlet–triplet gap sign classification. Its ability to reproduce all reference negative-gap systems while introducing relatively few false positives supports its selection for investigating the remaining molecular datasets.

\begin{figure*}[!htpb]
    \centering
    \includegraphics[width=\linewidth]{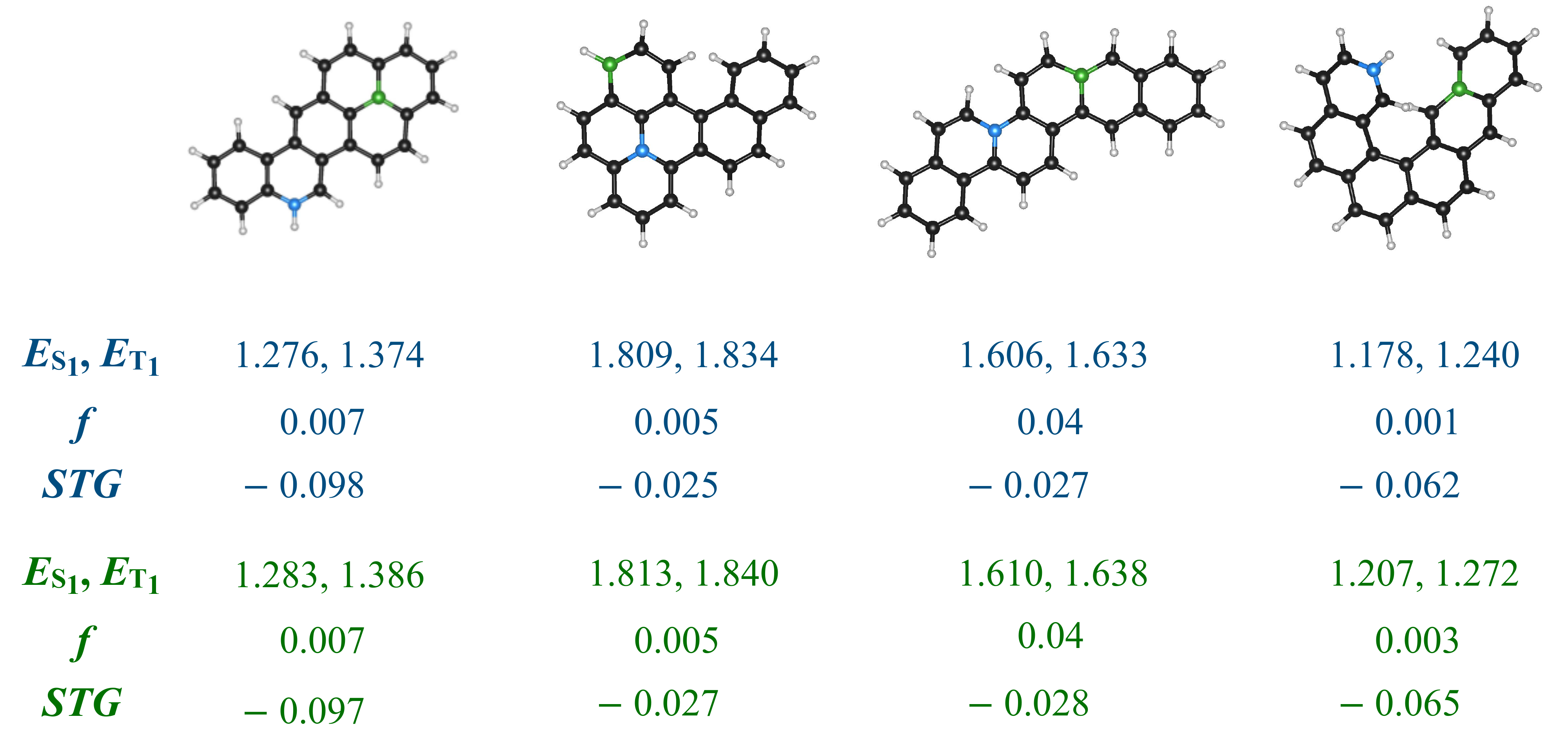}
    \caption{
Comparison of S$_1$ and T$_1$ excitation energies, singlet--triplet gaps (STGs), and S$_0$ $\rightarrow$ S$_1$ oscillator strengths (dimensionless) for four representative B,N-substituted polycyclic aromatic hydrocarbons exhibiting the most negative singlet–triplet gaps. 
We calculated all excited-state properties at the L-CC2/aug-cc-pVDZ level using geometries optimized with $\omega$B97X-D3 (blue) and $\omega$B97M-D4 (green). All energies are reported in eV.
}
\label{fig:BNPAH}
\end{figure*}

\subsubsection{B,N-substituted polyaromatic hydrocarbons}

The second dataset comprises 46 B,N-substituted polycyclic aromatic hydrocarbons (BNPAHs) previously identified as INVEST candidates from a screening of over 30,000 molecules.\cite{majumdar2025unlocking} In the original study, geometries were optimized at the $\omega$B97X-D3/def2-TZVP level, and singlet--triplet gaps were calculated using L-CC2/aug-cc-pVDZ. All 46 molecules exhibited negative singlet–triplet gaps with magnitudes exceeding 0.015 eV, the estimated uncertainty of the L-CC2 calculations.

To examine the sensitivity of these predictions to the geometry optimization method, we reoptimized all 46 structures at the $\omega$B97M-D4/def2-TZVP level and recalculated their excitation energies using L-CC2/aug-cc-pVDZ. Four optimized structures exhibited small imaginary vibrational frequencies below 1 THz in magnitude ($1~\mathrm{THz} = 33.356~\mathrm{cm}^{-1}$), which we retained for subsequent calculations. All 46 molecules prominently retained their negative singlet–triplet gaps at the $\omega$B97M-D4 geometries, supporting the validity of the previously reported INVEST predictions and indicating that they are not artifacts of the original geometry optimization method. Figure~\ref{fig:BNPAH} compares excitation energies and singlet–triplet gaps obtained with the two sets of optimized geometries for representative molecules.

\subsubsection{Non-alternant hydrocarbons}

The third dataset comprises 187 non-alternant hydrocarbons (NAHs) containing up to four fused five-, six-, and seven-membered rings. 
In the original study by Garner \textit{et al.},\cite{garner2023double} they optimized geometries at the $\omega$B97XD/def2-TZVP level and calculated excitation energies with CC2/aug-cc-pVDZ, yielding eight molecules with negative singlet--triplet gaps. 
Following the treatment adopted for the BNPAH dataset, we retained structures exhibiting imaginary vibrational frequencies of magnitude for further analysis.

\begin{figure}[!htpb]
    \centering
    \includegraphics[width=\linewidth]{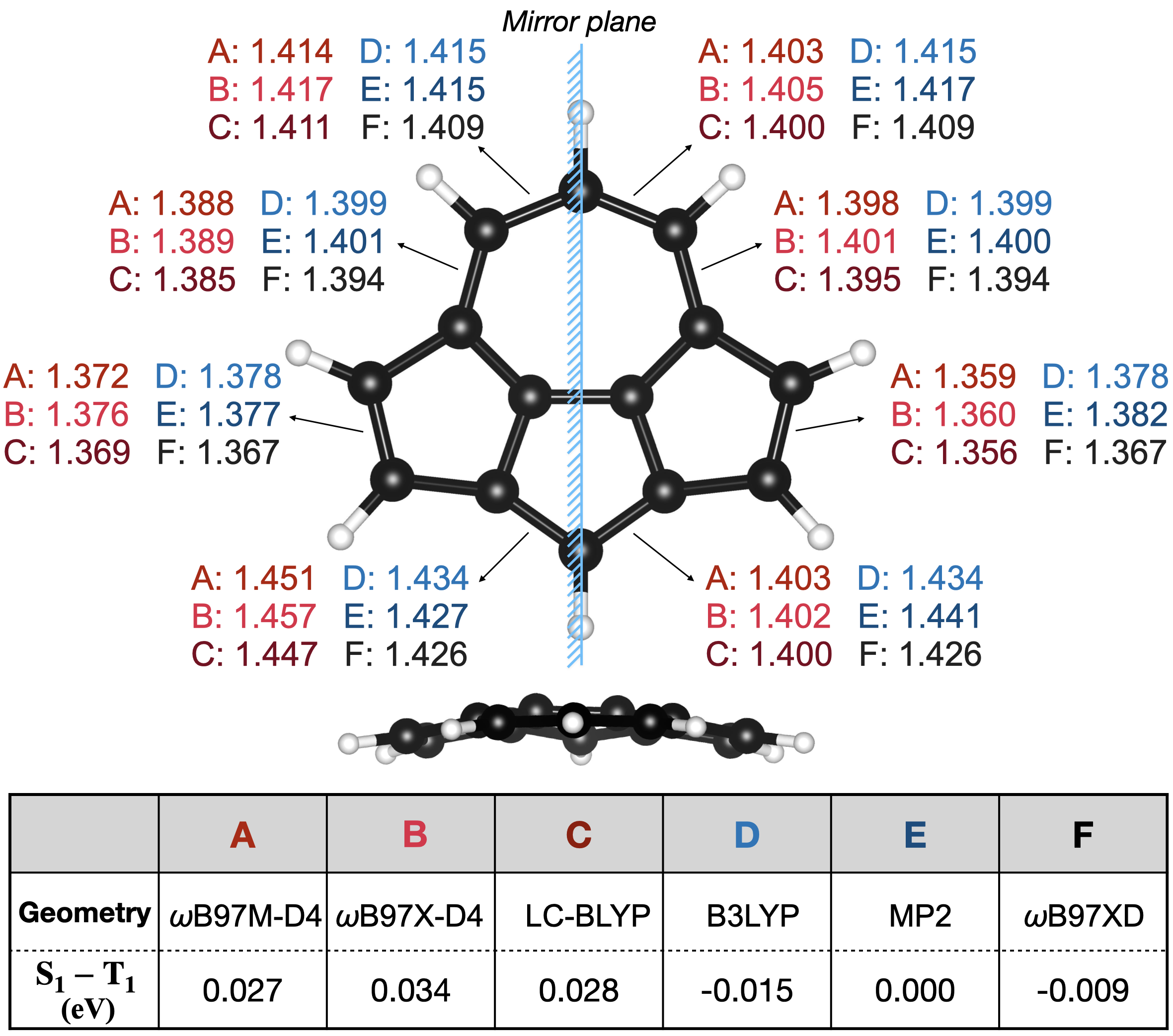}
    \caption{
   Molecular structure of a non-alternant hydrocarbon for which the choice of geometry optimization method qualitatively changes the predicted singlet–triplet gap from negative to positive.
   Bond lengths (in \AA) from six geometry optimization methods, A--F are shown, and the corresponding singlet–triplet gaps (in eV) calculated at the L-CC2/aug-cc-pVDZ level are also listed in the table below.
    }
\label{fig:mol35nah}
\end{figure}

In this work, we reoptimized all 187 molecules at the $\omega$B97M-D4/def2-TZVP level, followed by L-CC2/aug-cc-pVDZ excited-state calculations. 
Seven molecules were predicted to exhibit negative singlet–triplet gaps, compared with eight in the original study. 
The discrepancy arises from a molecule whose predicted singlet--triplet ordering depends on the geometry optimization method (Fig.~\ref{fig:mol35nah}).

\begin{table*}[!htpb]
\centering
\caption{
S$_1$ and T$_1$ excitation energies and singlet–triplet gaps for the 85 molecules predicted to exhibit inverted singlet–triplet (INVEST) gaps across the three datasets. All values were calculated at the L-CC2/aug-cc-pVDZ level using geometries optimized with $\omega$B97M-D4. The S$_0$ $\rightarrow$ S$_1$ oscillator strengths ($f$, in atomic units) are given in parentheses alongside the S$_1$ excitation energies.
}
\begin{tabular}{lccc c lccc}
\hline
INVEST candidates & S$_1$ (\textit{f}) & T$_1$ & gap && INVEST candidates & S$_1$ (\textit{f}) & T$_1$ & gap \\
\hline

INVEST-AP1 & $2.767$  (0.000) & $2.986$ & $-0.219$ && INVEST-BNPAH12 & $1.697$  (0.012) & $1.762$ & $-0.065$ \\
INVEST-AP2 & $2.164$  (0.000) & $2.332$ & $-0.168$ && INVEST-BNPAH13 & $1.207$  (0.003) & $1.272$ & $-0.065$ \\
INVEST-AP3 & $2.505$  (0.002) & $2.661$ & $-0.156$ && INVEST-BNPAH14 & $1.812$  (0.013) & $1.870$ & $-0.058$ \\
INVEST-AP4 & $1.599$  (0.000) & $1.735$ & $-0.136$ && INVEST-BNPAH15 & $1.307$  (0.002) & $1.361$ & $-0.054$ \\
INVEST-AP5 & $1.973$  (0.000) & $2.101$ & $-0.128$ && INVEST-BNPAH16 & $1.930$  (0.008) & $1.984$ & $-0.054$ \\
INVEST-AP6 & $1.066$  (0.000) & $1.193$ & $-0.127$ && INVEST-BNPAH17 & $1.084$  (0.002) & $1.138$ & $-0.054$ \\
INVEST-AP7 & $2.231$  (0.002) & $2.357$ & $-0.126$ && INVEST-BNPAH18 & $1.891$  (0.010) & $1.943$ & $-0.052$ \\
INVEST-AP8 & $2.166$  (0.006) & $2.289$ & $-0.123$ && INVEST-BNPAH19 & $1.327$  (0.000) & $1.376$ & $-0.049$ \\
INVEST-AP9 & $1.835$  (0.001) & $1.949$ & $-0.114$ && INVEST-BNPAH20 & $1.482$  (0.005) & $1.528$ & $-0.046$ \\
INVEST-AP10 & $1.805$  (0.001) & $1.917$ & $-0.112$ && INVEST-BNPAH21 & $1.268$  (0.000) & $1.314$ & $-0.046$ \\
INVEST-AP11 & $1.680$  (0.002) & $1.790$ & $-0.110$ && INVEST-BNPAH22 & $1.369$  (0.006) & $1.413$ & $-0.044$ \\
INVEST-AP12 & $1.990$  (0.001) & $2.098$ & $-0.108$ && INVEST-BNPAH23 & $1.218$  (0.002) & $1.261$ & $-0.043$ \\
INVEST-AP13 & $1.472$  (0.000) & $1.580$ & $-0.108$ && INVEST-BNPAH24 & $1.329$  (0.000) & $1.371$ & $-0.042$ \\
INVEST-AP14 & $1.619$  (0.003) & $1.725$ & $-0.106$ && INVEST-BNPAH25 & $1.578$  (0.014) & $1.619$ & $-0.041$ \\
INVEST-AP15 & $1.938$  (0.003) & $2.035$ & $-0.097$ && INVEST-BNPAH26 & $1.036$  (0.006) & $1.072$ & $-0.036$ \\
INVEST-AP16 & $1.470$  (0.000) & $1.562$ & $-0.092$ && INVEST-BNPAH27 & $1.588$  (0.005) & $1.617$ & $-0.029$ \\
INVEST-AP17 & $1.886$  (0.006) & $1.972$ & $-0.086$ && INVEST-BNPAH28 & $1.068$  (0.004) & $1.096$ & $-0.028$ \\
INVEST-AP18 & $2.050$  (0.002) & $2.134$ & $-0.084$ && INVEST-BNPAH29 & $1.610$  (0.039) & $1.638$ & $-0.028$ \\
INVEST-AP19 & $2.242$  (0.009) & $2.320$ & $-0.078$ && INVEST-BNPAH30 & $1.813$  (0.005) & $1.840$ & $-0.027$ \\
INVEST-AP20 & $2.031$  (0.001) & $2.102$ & $-0.071$ && INVEST-BNPAH31 & $1.328$  (0.001) & $1.352$ & $-0.024$ \\
INVEST-AP21 & $2.350$  (0.003) & $2.421$ & $-0.071$ && INVEST-BNPAH32 & $1.214$  (0.011) & $1.238$ & $-0.024$ \\
INVEST-AP22 & $2.011$  (0.005) & $2.073$ & $-0.062$ && INVEST-BNPAH33 & $1.828$  (0.007) & $1.851$ & $-0.023$ \\
INVEST-AP23 & $1.677$  (0.004) & $1.735$ & $-0.058$ && INVEST-BNPAH34 & $1.057$  (0.011) & $1.080$ & $-0.023$ \\
INVEST-AP24 & $1.831$  (0.003) & $1.883$ & $-0.052$ && INVEST-BNPAH35 & $1.418$  (0.015) & $1.440$ & $-0.022$ \\
INVEST-AP25 & $1.365$  (0.000) & $1.416$ & $-0.051$ && INVEST-BNPAH36 & $1.375$  (0.024) & $1.397$ & $-0.022$ \\
INVEST-AP26 & $2.354$  (0.001) & $2.401$ & $-0.047$ && INVEST-BNPAH37 & $1.343$  (0.031) & $1.363$ & $-0.020$ \\
INVEST-AP27 & $2.088$  (0.007) & $2.126$ & $-0.038$ && INVEST-BNPAH38 & $1.982$  (0.003) & $2.001$ & $-0.019$ \\
INVEST-AP28 & $1.590$  (0.004) & $1.621$ & $-0.031$ && INVEST-BNPAH39 & $1.088$  (0.009) & $1.107$ & $-0.019$ \\
INVEST-AP29 & $1.454$  (0.008) & $1.478$ & $-0.024$ && INVEST-BNPAH40 & $1.273$  (0.002) & $1.292$ & $-0.019$ \\
INVEST-AP30 & $1.419$  (0.005) & $1.437$ & $-0.018$ && INVEST-BNPAH41 & $1.827$  (0.027) & $1.845$ & $-0.018$ \\
INVEST-AP31 & $1.953$  (0.002) & $1.969$ & $-0.016$ && INVEST-BNPAH42 & $1.841$  (0.026) & $1.859$ & $-0.018$ \\
INVEST-AP32 & $1.952$  (0.002) & $1.964$ & $-0.012$ && INVEST-BNPAH43 & $2.047$  (0.012) & $2.065$ & $-0.018$ \\
INVEST-BNPAH1 & $1.283$  (0.007) & $1.380$ & $-0.097$ && INVEST-BNPAH44 & $1.222$  (0.012) & $1.239$ & $-0.017$ \\
INVEST-BNPAH2 & $1.390$  (0.002) & $1.474$ & $-0.084$ && INVEST-BNPAH45 & $1.155$  (0.005) & $1.172$ & $-0.017$ \\
INVEST-BNPAH3 & $1.850$  (0.012) & $1.926$ & $-0.076$ && INVEST-BNPAH46 & $2.202$  (0.006) & $2.217$ & $-0.015$ \\
INVEST-BNPAH4 & $1.758$  (0.013) & $1.830$ & $-0.072$ && INVEST-NAH1 & $1.989$  (0.000) & $2.122$ & $-0.133$ \\
INVEST-BNPAH5 & $1.072$  (0.001) & $1.142$ & $-0.070$ && INVEST-NAH2 & $1.881$  (0.000) & $1.937$ & $-0.056$ \\
INVEST-BNPAH6 & $1.921$  (0.012) & $1.990$ & $-0.069$ && INVEST-NAH3 & $1.537$  (0.007) & $1.556$ & $-0.019$ \\
INVEST-BNPAH7 & $1.604$  (0.004) & $1.672$ & $-0.068$ && INVEST-NAH4 & $1.817$  (0.000) & $1.831$ & $-0.014$ \\
INVEST-BNPAH8 & $1.383$  (0.004) & $1.451$ & $-0.068$ && INVEST-NAH5 & $0.958$  (0.003) & $0.971$ & $-0.013$ \\
INVEST-BNPAH9 & $1.846$  (0.005) & $1.914$ & $-0.068$ && INVEST-NAH6 & $1.496$  (0.000) & $1.502$ & $-0.006$ \\
INVEST-BNPAH10 & $1.305$  (0.009) & $1.371$ & $-0.066$ && INVEST-NAH7 & $2.300$  (0.007) & $2.306$ & $-0.006$ \\
INVEST-BNPAH11 & $1.686$  (0.005) & $1.752$ & $-0.066$ &&  \\
\hline
\end{tabular}
\label{tab:INVEST85}
\end{table*}

To examine the structural origin of this difference, we compared geometries obtained using six optimization methods.  
B3LYP and $\omega$B97XD preserve a molecular mirror-plane symmetry
corresponding to the \cs symmetry, whereas LC-BLYP, $\omega$B97X-D4, and $\omega$B97M-D4 lead to symmetry breaking (\cs $\rightarrow C_1$), accompanied by pronounced bond-length alternation (Fig.~\ref{fig:mol35nah}). 
The corresponding singlet–triplet gaps, evaluated consistently at the L-CC2/aug-cc-pVDZ level, reveal a qualitative dependence on geometry: the B3LYP and $\omega$B97XD geometries yield negative gaps, whereas the LC-BLYP, $\omega$B97X-D4, and $\omega$B97M-D4 geometries give positive gaps. 
The MP2 geometry yields nearly degenerate S$_1$ and T$_1$ states.

For this molecule, Garner \textit{et al.}\cite{garner2023double} reported a singlet–triplet gap of $-0.005$ eV at the CC2/aug-cc-pVDZ level using the $\omega$B97XD-optimized geometry. 
Our L-CC2/aug-cc-pVDZ calculation at the same geometry gives S$_1$ and T$_1$ excitation energies of 0.962 and 0.971 eV, respectively, corresponding to a singlet–triplet gap of $-0.009$ eV. 
In contrast, the $\omega$B97M-D4 geometry yields excitation energies of 1.052 and 1.025 eV, respectively, resulting in a positive singlet–triplet gap of $+0.027$ eV. 
Thus, changing the optimized geometry increases the singlet–triplet gap by 0.036 eV, reversing the predicted excited-state ordering.

These results demonstrate that the negative singlet–triplet gap previously predicted\cite{garner2023double} at the higher-symmetry geometry does not persist at the $\omega$B97M-D4-optimized structure. 
The associated symmetry lowering and bond-length alternation highlight the sensitivity of INVEST predictions to structural relaxation, consistent with earlier observations for distortion-prone molecular systems.\cite{majumdar2024influence}

\subsubsection{INVEST candidates}

Using $\omega$B97M-D4-optimized geometries and L-CC2/aug-cc-pVDZ excitation energies, we predicted negative singlet--triplet gaps for 85 of the 337 molecules investigated. 
These comprise 32 APs, 46 B,N-substituted polycyclic aromatic hydrocarbons (BNPAHs), and 7 non-alternant hydrocarbons (NAHs). The S$_1$ and T$_1$ excitation energies, corresponding singlet–triplet gaps, and S$_0$ $\rightarrow$ S$_1$ oscillator strengths for all 85 candidates are summarized in Table~\ref{tab:INVEST85}. 
Among the three molecular families, APs exhibit the largest negative singlet–triplet gaps, with heptazine (INVEST-AP1) having the most negative gap across the entire dataset.

The oscillator strengths reveal distinct characteristics among the molecular families. All 32 APs exhibit negligible or vanishing S$_0$ $\rightarrow$ S$_1$ oscillator strengths, consistent with the weakly allowed or symmetry-forbidden transitions. 
In contrast, several BNPAHs show appreciable oscillator strengths while retaining negative singlet–triplet gaps, highlighting their potential to combine INVEST ordering with optically allowed transitions.

%=========================================
\section{Conclusions\label{sec:conclusion}}
%=========================================
The choice of geometry optimization method plays an essential role in reliably predicting INVEST character, particularly in molecules susceptible to symmetry-breaking distortions. 
Our results demonstrate that commonly used methods such as B3LYP and MP2 can favor artificially high-symmetry geometries, leading to substantial errors in singlet--triplet gaps and, in some cases, false-positive INVEST assignments. Further, 
accurate prediction of the molecular symmetry alone is insufficient, as the magnitude of the structural distortion can also significantly influence the excited-state ordering.

By assessing geometry optimization methods for 13 pseudo--Jahn--Teller distortion-prone azaphenalenes against CCSD(T)-level reference geometries, we identified $\omega$B97M-D4 as a suitable compromise between accuracy and computational efficiency. 
We further tested its reliability on 104 azaphenalenes, reproducing all 30 reference negative-gap systems with only two false positives. 
In contrast, B3LYP and MP2 produced 31 and 29 false positives, respectively, demonstrating the consequences of an inadequate description of symmetry-breaking distortions.

We then applied $\omega$B97M-D4 to optimize the geometries of 337 molecules spanning three molecular families, with excited-state energies calculated consistently at the L-CC2/aug-cc-pVDZ level. 
We report negative singlet--triplet gaps for 32 azaphenalenes, 46 B,N-substituted polycyclic aromatic hydrocarbons, and 7 non-alternant hydrocarbons, giving a total of 85 candidates. 
All 46 previously reported BNPAH negative-gap predictions were retained, supporting their robustness across the geometry optimization methods examined. 
In contrast, one non-alternant hydrocarbon previously predicted to exhibit a negative singlet--triplet gap yielded a positive gap at the $\omega$B97M-D4 geometry, highlighting how symmetry lowering and bond-length alternation can qualitatively alter the predicted excited-state ordering. The azaphenalenes generally exhibit negligible S$_0$ $\rightarrow$ S$_1$ oscillator strengths, whereas some BNPAHs combine negative singlet--triplet gaps with appreciable oscillator strengths, indicating potentially more favorable optical transitions. 

Overall, both the ability to capture symmetry breaking and the accuracy of the resulting distorted geometry are essential for reliable computational INVEST screening. 
For the molecular families examined, $\omega$B97M-D4 offers a computationally practical alternative to higher-level geometry optimization, reducing the risk of spurious INVEST assignments while enabling the investigation of larger conjugated systems.

\section{Supplementary Information}
Table S1: L-CC2/aug-cc-pVDZ excitation energies of 13 benchmark systems.
Table S2: Excitation energies and singlet–triplet gaps of 30 APs at CCSD(T) and DFT geometries.
Geometries: Cartesian coordinates and harmonic frequencies of 85 INVEST candidates at the $\omega$B97M-D4/def2-TZVP level.
The INVEST85 dataset~\cite{INVEST85} contains the optimized molecular geometries and L-CC2 excitation energies for all systems investigated in this work.

\section{Data Availability}
The data supporting the findings of this study are provided in the article and its supplementary material.

\section{Acknowledgments}
We acknowledge the support of the 
Department of Atomic Energy, Government
of India, under Project Identification No.~RTI~4007. 
All calculations have been performed using the Helios computer cluster, 
which is an integral part of the MolDis 
Big Data facility, 
TIFR Hyderabad \href{http://moldis.tifrh.res.in}{(http://moldis.tifrh.res.in)}.

\section{Author Declarations}

\subsection{Author contributions}
\noindent 
{\bf AM}: 
Conceptualization (equal); 
Analysis (equal); 
Data collection (equal); 
Review \& editing (equal);
Writing (equal).
{\bf ESS}:
Analysis (equal); 
Review \& editing (equal).
{\bf RR}: Conceptualization (equal); 
Analysis (equal); 
Data collection (equal); 
Funding acquisition; 
Project administration and supervision; 
Resources; 
Review \& editing (equal);
Writing (equal).

\subsection{Conflicts of Interest}
The authors have no conflicts of interest to disclose.
%\singlespacing

%\clearpage
\section{References}
\bibliography{ref} 
\end{document}